\documentclass[aps,twocolumn,superscriptaddress,showpacs,floatfix]{revtex4}
\usepackage{amsmath,epsfig,subfigure}
\usepackage{tabularx,multirow}
\usepackage[active]{srcltx}

\begin{document}
\title{Energetics and dynamics of H$_2$ adsorbed in a nanoporous material at low temperature}
\author{Lingzhu Kong}
\affiliation{Department of Physics and Astronomy, Rutgers University, Piscataway, New Jersey 08854-8019, USA}
\author{Guillermo Rom\'{a}n-P\'{e}rez}
\affiliation{Departamento de F\'isica de la Materia Condensada, C-III, Universidad Aut\'{o}noma de Madrid, E-28049 Madrid, Spain}
\author{Jos\'{e} M. Soler}
\affiliation{Departamento de F\'isica de la Materia Condensada, C-III, Universidad Aut\'{o}noma de Madrid, E-28049 Madrid, Spain}
\author{David C. Langreth}
\affiliation{Department of Physics and Astronomy, Rutgers University, Piscataway, New Jersey 08854-8019, USA}

\begin{abstract}
Molecular hydrogen adsorption in a nanoporous metal organic framework structure (MOF-74) 
was studied via van der Waals density-functional calculations. 
The primary and secondary binding sites for H$_2$ were confirmed. The low-lying 
rotational and translational
energy levels were calculated, based on the orientation and position dependent
potential energy surface at the two binding sites. A consistent picture 
is obtained between the calculated rotational-translational transitions 
for different H$_2$ loadings and 
those measured by inelastic neutron scattering exciting the singlet 
to triplet (para to ortho) transition in H$_2$.
The H$_2$ binding energy after zero point
energy correction due to the rotational and translational motions is predicted
to be $\sim$100 meV in good agreement with the experimental value of $\sim$90 meV.
\end{abstract}

\maketitle

The adsorption of molecules within the nanopores of a sparse material and
their low-lying excitations provide rich phenomena of fundamental interest
that are seldom explored by first
principles methods, even for molecules as simple as H$_2$. For example, one
can ask how the well-known para--ortho transition of H$_2$ survives in a recognizable
form in the presence of the rotational barriers and hindrances provided by the
adsorbing material, and how it changes for different concentrations of H$_2$.
The necessity of proper treatment of van der Waals interactions has foiled
traditional density functional treatments.  Here, by using the van der Waals
density functional (vdW-DF) of Dion et al.~\cite{vdw}, we are able to predict a picture of
the low-lying excitations which provides a credible match for the results
of inelastic neutron diffraction (INS)~\cite{Liu2008} for different hydrogen loadings
in a nanoporous material of a type thought to provide a possible path toward
future hydrogen storage technology.

The material studied here is a metal-organic framework (MOF) compound,
namely the material that has been called MOF-74~\cite{Yaghi2005},
chosen because of recent inelastic neutron scattering measurements~\cite{Liu2008}
with varying amounts of H$_2$ adsorption.
More generally MOFs are
a relatively new class of materials
which are composed of metal clusters connected by organic 
ligands~\cite{Eddaoudi2001}.
The search for a MOF structure with high binding strength 
to H$_2$ has become very active recently~\cite{Zhou2008,Vitillo2008}. 
The dynamical properties of the adsorbed dihydrogen such as vibrational
and hindered rotational motion play a significant role in determining
the H$_2$ binding energy, especially for caged structures with nanopores.
This area is even less explored than the modeling of hydrogen uptake based
only on the depth of the potential well. An attempt along this line
studied the rotational transition of H$_2$ adsorbed over benzene molecule
which is often a fraction of the organic linker in MOF 
materials~\cite{Hamel2004}. However, the interaction between full MOF 
and H$_2$ usually differs significantly 
from that between a fraction of MOF and H$_2$~\cite{Kong2009}.
A very recent paper studied the rotational transition in another MOF (HKUST-1) with 
generalized gradient approximation (GGA) calculations~\cite{Brown2009}. 
However, GGA is known to fail for
cases where the binding is dominated by van der Waals interactions,
and only a preliminary theoretical picture was obtained.

Here using 
vdW-DF~\cite{vdw},
implemented via the efficient algorithm of two of us~\cite{Soler2009short},
we calculated the dynamical properties as well as the binding energy of H$_2$ 
adsorbed in MOF-74 structure, which has been demonstrated to increase both 
the H$_2$ affinity~\cite{Zhou2008} and hydrogen density~\cite{Liu2008} 
due to the unsaturated metal sites~\cite{Yaghi2005}. The primary
binding sites associated with metal atoms are confirmed by our
calculations with a binding energy of 130 meV, while the
LDA and GGA results are 230 and 46 meV respectively~\cite{Zhou2008}.
The center-of-mass translation and molecular rotation of H$_2$ 
are found to attribute zero point energy (ZPE) of $\sim$20 meV 
and $\sim$10 meV, respectively,  to the total energy of the MOF--H$_2$ system. 
As a result the effective binding is about 100 meV, which is close to
the experimentally measured value of 
$\sim$90 meV~\cite{Liu2008,Yaghi2005}.

Furthermore, we find that the adsorbed H$_2$ is a three-dimensional
quantum rotor at both the primary and secondary binding sites.
The original $J=1$ triplet state splits into three nondegenerate levels,
 which imply three 
distinct para to ortho excitation energies.
A good
match between the calculated rotational transitions and the inelastic 
neutron scattering spectra~\cite{Liu2008} was obtained.
As suggested in 
Ref.~\onlinecite{Yildirim2005}, MOF materials may be used to create
artificial hydrogen nanocages. In addition to the confinement and short
intermolecular distance, the temperature dependent preferential adsorbance \cite{Silvera1976}
 of ortho-H$_2$
due to the splitting of the triplet could give rise to unexpected properties
as well.

\begin{figure}[t]
{\epsfig{file=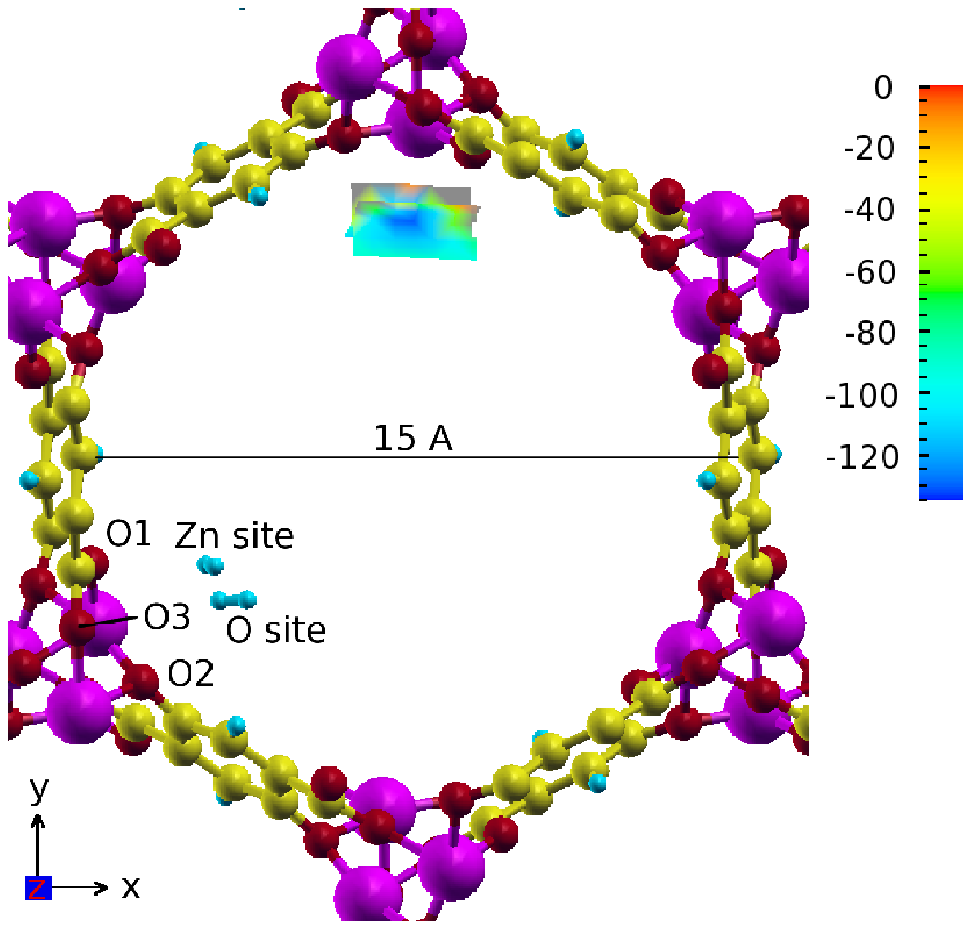,width=2.5in,clip=true} \hspace{.5in} 
	\epsfig{file=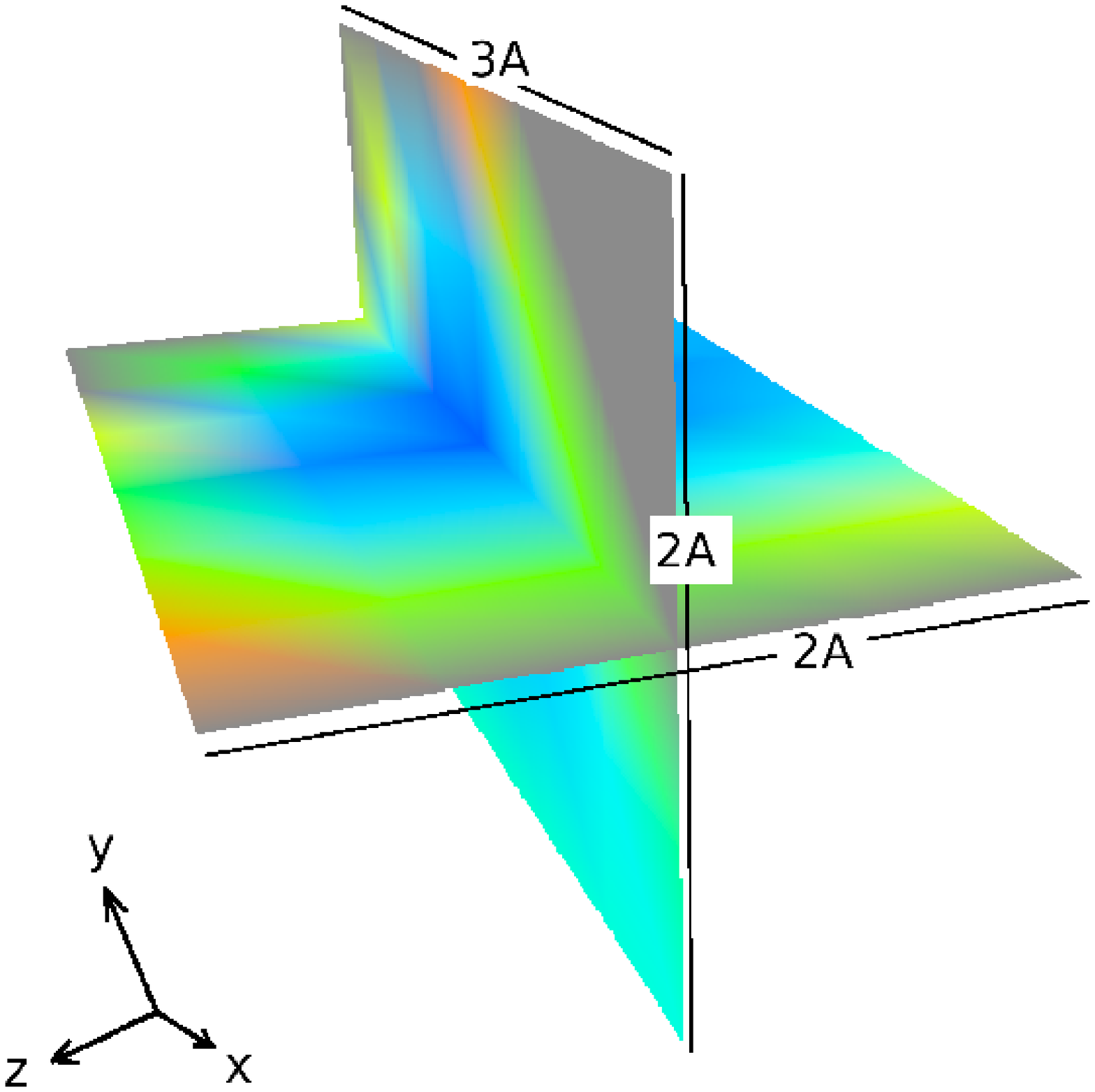,width=2.4in,clip=true}}
\caption{ Potential energy surface at the Zn site (meV). Top panel:
cell with energy map  plus the location of primary and secondary sites at an equivalent
place in the cell; the two sites have different
depths.
Bottom panel: blowup of energy map in a different orientation.}
\label{fig:ZnsitePES}
%
\end{figure}

To locate the binding site near the Zn atom, we used the
experimental MOF
structure. With one H$_2$ per primitive cell (1/6 H$_2$ per Zn), we calculated the 
MOF--H$_2$ interaction energy for various H$_2$ positions near the Zn atom.
The contour map of the interaction energy is shown in 
Fig.~\ref{fig:ZnsitePES}. The minimum position has a binding energy of 
130 meV and its distance to the Zn atom is 2.9 \AA~while the 
experimental value is 2.6 \AA~\cite{Liu2008}. The over-estimation of 
the bond length is typical in vdW-DF calculations~\cite{Langreth2009}.

Along with the strongest binding site
near the metal atom
(hereinafter referred to as the Zn site), a secondary site
above the oxygen triangle was observed by neutron scattering 
measurements~\cite{Liu2008}. We again took the structure of 
Ref.~\onlinecite{Liu2008} with a loading of
12 D$_2$/cell. Instead of sampling the potential energy surface 
for the two H$_2$ manually, we performed a relaxation of the adsorbed
dihydrogens with a fixed MOF host with the force converged to 12 meV/\AA.
We find that a H$_2$ indeed positions itself above the oxygen triangle 
when its neighboring metal sites are occupied. In other words, 
a potential minimum is created at the O-triangle site due to the
adsorbance of H$_2$ at the metal site. The closest distance between sorbate and sorbent 
is 3.4~\AA{}, while the distance between the O and Zn sites is  3.2~\AA{}.  These
are each 0.3~\AA{} larger than the respective experimental distances~\cite{Liu2008}.

With the two binding sites located, we now turn to the dynamical properties
of the adsorbed H$_2$ at these sites, which are important in determining 
the zero-point energy correction to the total binding strength. 
In general, H$_2$ has six degrees of freedom. We here focus on the 
translational and rotational motions where the H$_2$ bond length is
fixed at its equilibrium value.
Even with fixed bond length, the
full quantum mechanical treatment of the remaining five degrees of freedom
is still beyond the scope of current letter. 
We make the approximation that rotation-translation 
coupling may be
neglected to lowest order, so that the two motions can be treated 
separately. In the absence of intersite dynamical coupling, this would
be exact to the extent that the H$_2$ motions were small enough to be described by
the harmonic approximation.  Neither of these is exact, so that small
deviations from the rotational spectra we calculate are to be expected,
even though we use methods that do not explicitly assume the harmonic
approximation's validity.

\begin{figure}[h]
\centerline{\epsfig{file=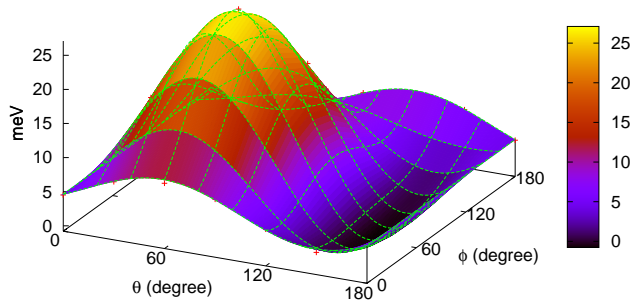,width=3.4in,clip=true}}
\caption{ Orientational dependence of the potential at the Zn site. The spherical coordinates
are defined with respect to the axes of Fig.~\ref{fig:ZnsitePES}.
The hydrogen loading is 1 H$_2$/cell.}
\label{orientation-Znsite}
\end{figure}

Fig.~\ref{orientation-Znsite} shows the orientational dependence of the
interaction energy at Zn site. 
The maximum point in this figure corresponds to H$_2$ bond 
aligned toward the Zn atom, which is
the weakest binding orientation, while the preferential bond direction of 
the adsorbed dihydrogen is in the plane perpendicular to the Zn--H axis.
The O site, however,
is quite different from the Zn site. The H$_2$ tends to line up toward 
the oxygen triangle and the preferential orientation is inside a conical
region (figure not shown). The maximum variation of the rotational potential at 
each of the two
sites is $\sim$27 meV and $\sim$18 meV, respectively. Due to this variation, the adsorbed
hydrogen molecule exhibits hindered rotation behavior at these two sites.
As a result, the energy levels of the free H$_2$ rotor ($E=BJ(J+1), B=7.35$ meV)
will be lifted and, in particular, three nondegenerate states emerge from 
what was the $J=1$ level with
$m=0, \pm 1$, respectively. The transition between the
singlet and triplet states
is the well-known para-ortho transition. 
A spin flip is required to satisfy the antisymmetry requirement 
of the total wave function.
To compare with
INS data~\cite{Liu2008}, we calculated the rotational spectrum by solving
the rigid rotor Schr\"{o}dinger equation for H$_2$. We first fitted the 
orientational potential $E(\theta,\phi)$ to a summation of spherical harmonics
and then diagonalized the Schr\"{o}dinger equation with a spherical harmonics
basis. The results are shown in Table~\ref{table:theory-exp}, where three spectra
with different loadings are calculated for Zn site and  one set 
 for the O site, which is mostly unoccupied at low loading.

\begin{figure}[b]
%
\epsfig{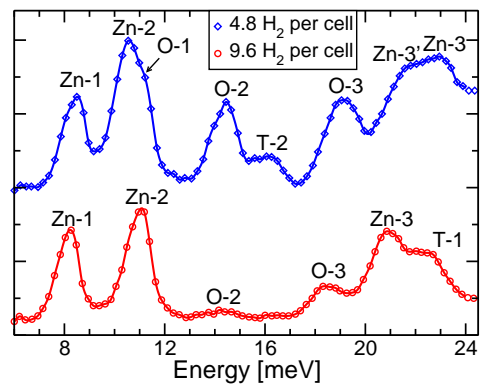}
\caption{ 
 Our assignments for the observed  energy loss peaks. The circles and diamonds are the 4 K inelastic neutron 
scattering data points of Ref.~\onlinecite{Liu2008} with the ordinate origins shifted arbitrarily.
 We put cubic splines through the data points as a guide to the eye.
}
\label{fig:liuINS}
\end{figure}

In Table~\ref{table:theory-exp} we first observe that the ground state energies
(zero point energies) are 
about 10 meV at both sites, and it increases by about 3 meV at Zn site as 
the loading is increased. 
The excited state of $J=1$ splits into three nondegenerate states for all four
cases considered theoretically, showing the hindered three-dimensional quantum rotor character of the 
adsorbed dihydrogen at the two sites. For the Zn site with 1 H$_2$/cell loading, the
three para-ortho transitions are 10.6, 14 and 21.6 meV, which we assign  to
 the main INS peaks at 8.3, 11.1, and 20.9 meV in Fig.~S2 of 
Ref.~\onlinecite{Liu2008}, whose points we replotted in Fig.~\ref{fig:liuINS}
for easy comparison. In addition to the three major peaks, two small 
peaks at 14.2 and 18.6 meV are also observed experimentally in the INS 
spectrum. Their strength increases substantially as the loading increases so that 
the second site begins to be occupied, and they move to slightly higher
energies (14.5 and 19.2 meV, respectively). This result is in surprisingly good 
agreement with our calculations which show that the O site gives two transitions 
at 15.6 and 18.5 meV. Moreover, the experimental peak at 11.1 meV significantly increases in
strength with a shoulder appearing under high H$_2$ loading; this feature
probably consists of two peaks with one of them  due to the second site  and the 
other due to the Zn site. This result is again consistent with our findings that
H$_2$ at the O site has a para-ortho transition at $\sim$11 meV. From Table~\ref{table:theory-exp}
we also see that the predicted transition energies at Zn site shift as loading increases,
especially for the Zn-3 transition, which changes from $\sim$22 to $\sim$26 meV while
the other two shift downward only by $\sim$1 meV.
Experimentally, there is a smaller corresponding shift
to about 23 meV for higher loading (9.6 H$_2$/cell).
For the  loadings of 7 and 9.6
H$_2$/cell,  we assign Zn-$3'$ to the contribution from the
sites where the neighboring O site is empty, while assigning Zn-3 to the 
case where it is occupied.  Theoretically there are no empty O sites at the 12 H$_2$/cell loading,
while presumably  at the 7 H$_2$/cell loading, 
none of the 
5  Zn-3 sites with empty O sites would 
show
the large shift that would be caused by a filled adjacent O site.

\begin{center}
\begin{table}[t]
\caption{Theory vs.\ experiment (Fig.~\ref{fig:liuINS}) for the positions (meV) of 
rotational energy-loss peaks along
with the calculated rotational zero point energy.
(RZPE).
}
\label{table:theory-exp}
\begin{tabular*}{0.48\textwidth}{@{\extracolsep{\fill}}ccccccc}
\hline
\hline
& \# H$_2$/cell &Method &RZPE &Zn-1 &Zn-2&Zn-3     \\
\hline
& 1	&theory		&10.1	&10.6 	&14.0 	&21.6  	\\
& 4.8	&experiment	&	&8.3	&11.1	&20.9	\\
& 7	&theory		&13.6	&9.2	&13.5	&26.2	\\
& 9.6	&experiment	&	&8.6	&10.6	&23.0	\\	
& 12 	&theory 	&12.9	&9.8	&12.8 	&26.4  	\\
\hline
&&&&O-1 &O-2&O-3     \\
\hline
&9.6	&experiment	&	&11.2	&14.5	&19.2	\\
& 12    &theory 	&9.6	&11.0 	&15.6 	&18.5 \\
\hline
\hline
\end{tabular*}
\end{table}
\end{center}

In Table~\ref{table:tran} we show, for different sites and loadings, the translational energies along
normal coordinate directions, obtained from the forces induced
by small displacements.
The energies were obtained by using these harmonic normal coordinates
to extend the results  approximately into the anharmonic quantum-mechanical
regime,
by first displacing the H$_2$ along these normal 
directions and then substituting the resulting potential
into
the  Schr\"{o}dinger equation to get the energies~\cite{Kong2009},
and hence the transition frequencies. The frequencies determined
by this method deviate as much as $\sim$10\% from the harmonic
frequencies, although for most cases the deviation is less than 5\%.
Table~\ref{table:tran} shows that the translational frequencies are 
remarkably close to the rotational transitions shown in Table~\ref{table:theory-exp} 
and could lead to a complex INS spectra. However, the scattering probability for
para-ortho transition is proportional to the incoherent cross 
section ($\sigma _i$) of atomic hydrogen while the para-para scattering 
is proportional to the coherent cross section ($\sigma _c$)~\cite{Young1964}. 
These cross sections have been tabulated~\cite{Sears1987} and one finds
that $\sigma_i \gtrsim 40 \sigma _c$.
Therefore the INS spectra would be 
dominated by the para-ortho transitions with little direct contribution from 
translational motions. 
Simultaneous excitation of rotational and translational
excitations  may appear as shoulders or 
sidebands of the major para-ortho transition peaks, 
which are not necessarily weak.
Although two possibilities are labeled (T-1 and T-2) in Fig.~\ref{fig:liuINS},
an unambiguous assignment is difficult. The majority of such double
excitations would be predicted to occur at frequencies higher than those
measured, or else hidden by the strong Zn-3 structures assigned to rotational
excitations. Spectral structure might also be produced by dynamical coupling
between hydrogen at adjacent sites.
A para-para 
sensitive tool such as Raman or IR may shed further light onto the origin 
of various structures.

\begin{center}
\begin{table}[t]
\caption{ Calculated translational frequencies (meV/$\hbar$) of adsorbed H$_2$.} \label{table:tran}
\begin{tabular*}{0.48\textwidth}{@{\extracolsep{\fill}}lccccc}
\hline
\hline

                        &\# H$_2$/cell    & n1    & n2   &  n3   \\
\hline
\multirow{2}{*}{Zn site}& 1               & 7.1 & 15.2  & 17.5   \\
                        & 12                 & 9.2 & 12.6  & 22.9   \\
\hline
               {O site }& 12                 & 9.5 & 14.5  & 21.7   \\

\hline \hline
\end{tabular*}
\end{table}
\end{center}

With the rotational and translational energy levels calculated, we
are able to obtain the zero point energy corrections to the total
binding strength, which are about 10 and 20 meV respectively.
Therefore, the effective binding strength at low loading (Zn site) is $\sim$100 meV.
It is somewhat larger than the experimental value of 
91~meV~\cite{Liu2008} (88 meV in Ref.~\onlinecite{Yaghi2005}). The slight
overestimation of the H$_2$ binding energy by vdW-DF is also observed for 
another MOF~\cite{Kong2009}. For the 
O site, our calculated binding including
zero point correction is $\sim$90 meV. This latter value would probably be
smaller relative to the prediction for the Zn site,
if it were not for a known inaccuracy of an extremal limit in the exchange
part of the functional~\cite{Eamonn09}. The experimental value~\cite{Liu2008}
is $\sim$50 meV.

\begin{figure}[t]
\begin{centering}
\epsfig{file=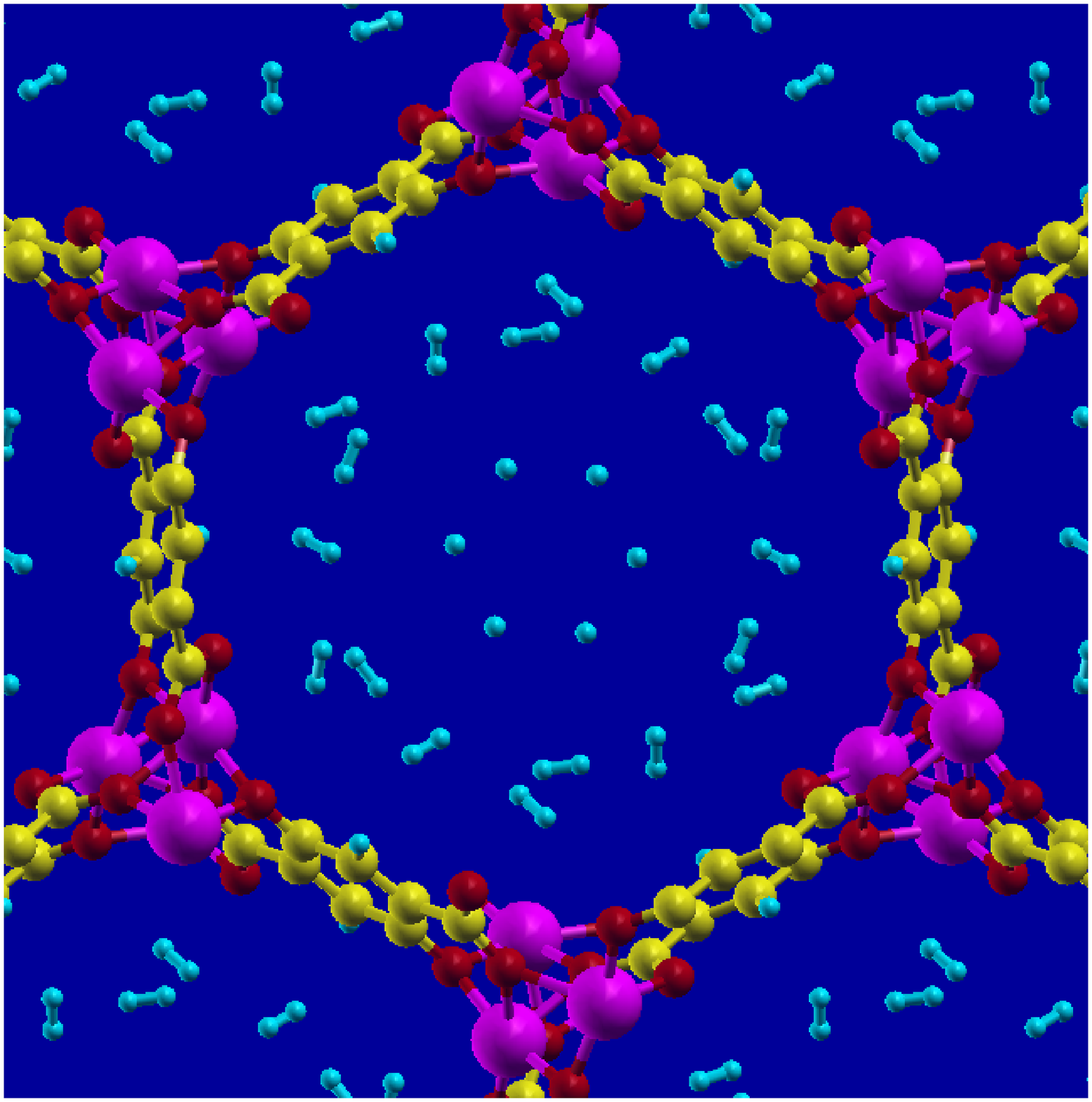,width=1.5in,clip=true}
\hspace{.2in}
\epsfig{file=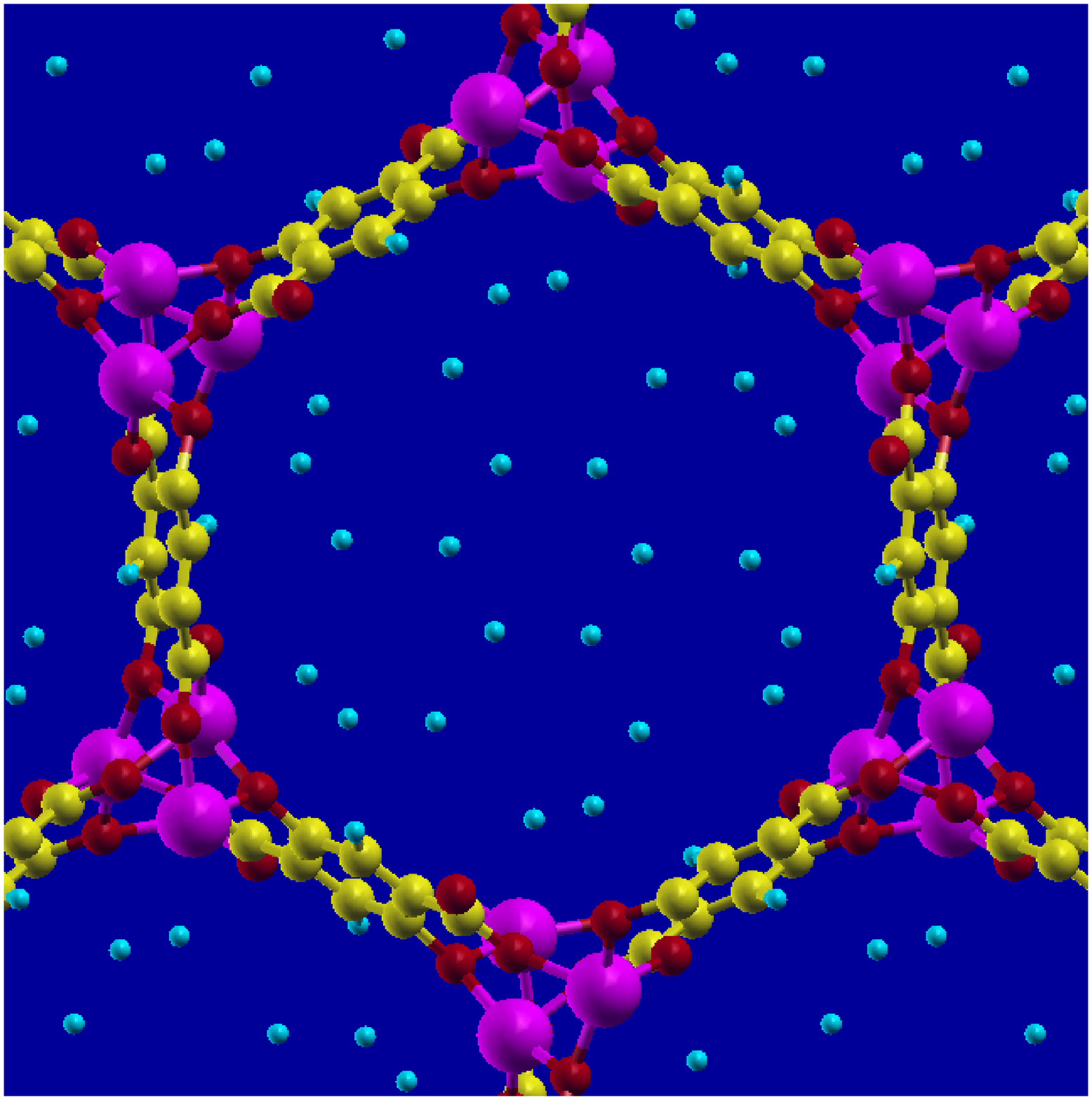,width=1.5in,clip=true}
\end{centering}
\caption{The adsorption sites. Left panel: our calculational results for H$_2$ or D$_2$
without consideration of zero point motions. 
Right panel: the neutron diffraction results~\cite{Liu2008} for D$_2$ at 4 K. The dihydrogens
shown are located at eight different distances perpendicular to the plane of the figure within
each primitive cell.}
\label{fig:24H2}
\end{figure}

Finally, we discuss the hydrogen uptake in this compound. As shown in 
Ref.~\onlinecite{Liu2008}, a total of 24 D$_2$/cell can be loaded into the
system to form a structure as shown in the right panel of Fig.~\ref{fig:24H2},
where each hydrogen molecule is represented by a single point.
To make a comparison, we put 24 H$_2$  per  cell
into the MOF and let them relax. 
To speed up the relaxation,  the dihydrogens were initially positioned 
to satisfy the 3-fold rotation plus inversion symmetry, with 6 near
each Zn site, 6 near each O site, 6 around 3 {\AA} from each aromatic
ring, and 6 at random in the central region. The final result is shown in the left panel of
Fig.~\ref{fig:24H2} where both hydrogen atoms in H$_2$ are displayed.
Overall agreement is obtained between theory and experiment
with the
the sorbate-sorbent distances in our calculations
deviating 
from experiment
by about 15\%.

In summary, we studied H$_2$ adsorbed 
within the nanopores of a complex metal organic framework
material (MOF-74)~\cite{Yaghi2005}
with vdW-DF calculations.  We confirmed the primary and secondary binding sites
associated with unsaturated metal atoms and   with triangles of O atoms, respectively.
The predicted (zero point corrected) binding energy of
the primary site is in good agreement with heat of adsorption measurements~\cite{Liu2008,Yaghi2005}
at low H$_2$ loading. We also identify a third and fourth type of site,
which are fully occupied at an H$_2$ concentration of 24 H$_2$ per 
primitive cell, in agreement with experiment~\cite{Liu2008}.
Rotational motion
of the adsorbed H$_2$ at the primary and secondary sites is analyzed.
The dihydrogen behaves as three-dimensional quantum rotor at both sites 
and the hindered rotation splits the triplet state into three nondegenerate levels. 
Based on the calculated para--ortho transitions from singlet to triplet, 
we are able to identify the main features in the INS spectra~\cite{Liu2008}  for different
H$_2$ concentrations, with agreement on the 20\% (5\%) level for the primary
(secondary) sites.

Supported by DOE Grant No.\ DE-FG02-08ER46491.

\bibliography{draft}

\end{document}